\documentclass[10pt]{article} 
\usepackage[utf8]{inputenc}
\usepackage{authblk}
\usepackage{tikz} 
\usepackage{tabularx}
\usepackage{appendix}

\usepackage[utf8]{inputenc}
\usepackage{appendix}

\usepackage[utf8]{inputenc}
\usepackage{authblk}
\usepackage{tikz} 
\usepackage{tabularx}
\usepackage{appendix}
\usepackage[english]{babel}
\usepackage{amsthm}
\usepackage{lineno}

\usepackage{shorthand}
\usepackage{amsmath}
\newcommand\norm[1]{\left\lVert#1\right\rVert}

\usepackage{biblatex} 
\title{On the mathematics of the circular flow of economic activity with applications to the topic of caring for the vulnerable during pandemics \\~\\}

\date{}

\author{ by \\

\textsc{Aziz Guergachi}\\

\textsc{Javid Hakim}\\

}

\begin{document}

\vspace{1.9cm}
\maketitle
\begin{abstract}

\vspace{0.3cm}

{\normalsize \noindent We investigate, at the fundamental level, the questions of `why', `when' and `how' one could or should reach out to poor and vulnerable people to support them in the absence of governmental institutions. We provide a simple and new approach that is rooted in linear algebra and basic graph theory to capture the dynamics of income circulation among economic agents. A new linear algebraic model for income circulation is introduced, based on which we are able to categorize societies as fragmented or cohesive. We show that, in the case of fragmented societies, convincing wealthy agents at the top of the social hierarchy to support the poor and vulnerable will be very difficult. We also highlight how linear-algebraic and simple graph-theoretic methods help explain, from a fundamental point of view, some of the mechanics of class struggle in fragmented societies. Then, we explain intuitively and prove mathematically why, in cohesive societies, wealthy agents at the top of the social hierarchy tend to benefit by supporting the vulnerable in their society. A number of new concepts emerge naturally from our mathematical analysis to describe the level of cohesiveness of the society, the number of degrees of separation in business (as opposed to social) networks, and the level of generosity of the overall economy, which all tend to affect the rate at which the top wealthy class recovers its support money back. In the discussion on future perspectives, the connections between the proposed matrix model and statistical physics concepts are highlighted.} 

\end{abstract}



\section{Introduction}

As the COVID-19 pandemic wreaked havoc around the world, most of the victims of this pandemic have been among the most vulnerable people in society,  ranging from  seniors in long-term care homes, homeless individuals and people with  underlying conditions such as  diabetes or hypertention. Most of these people who lost their lives to the pandemic in the USA  and Canada, at least, lived in poor neighborhoods. The pandemic has also accentuated socio-economic inequality, which has been on the rise over the last decades. According to a recent report by Oxfam, over 160 million people were forced into poverty due to the pandemic, while the number of billionaires in 2021 went up by almost 25\%.\\

In this paper, we propose an approach that is rooted in linear algebra and basic graph theory to help guide our thinking about how we can reach out to and protect the poor and vulnerable. More specifically, we make use of simple network scientific methods to capture the dynamics of income circulation among agents and provide insights about the situation of the poor and vulnerable in society. While doing that, we also attempt to address the following question: `why one should or shouldn't reach out to vulnerable people and support them?', which is, from a pure economic point of view, a legitimate question to study.\\

\section{The act of selling and buying as \textit{a} driver of money circulation}
A fundamental concept that underlies the notion of scarcity, which is the central core of economics (e.g.,\cite{mankiw}, page 4; \cite{samuelson}, page 4), is the fact that there are buyers who desire to acquire the scarce goods and services in the market \cite{PengAziz}. If no one is interested in these goods and services, then they would be irrelevant to the study of economics, even if they were scarce. The approach we propose in this paper to address the question of caring for the vulnerable focuses on the very exchange of goods and services between sellers and buyers, and attempts to capture the dynamics of this exchange from the ground up.\\

 To describe the basic idea of this approach, let us imagine a hypothetical community of three economic agents, who sell goods and services to each other in exchange for money, all in a closed economy. We refer to these three agents as A, B and C, also known as Alice, Bob and Carol. As a result of the business exchange taking place among them, Alice, Bob and Carol earn income. Let us denote the incomes earned by Alice, Bob and Carol up until time $t$ and expressed in monetary units as $x_1(t)$, $x_2(t)$ and $x_3(t)$ respectively. Between time $t$ and time $t+1$, Alice will sell goods and services to Bob who will then pay Alice a certain price $p$ for these goods and services. This price $p$ paid by Bob to Alice will come out of Bob's income $x_2(t)$ at time $t$; it will be a fraction of $x_2(t)$. Let us denote this fraction of $x_2(t)$ paid by Bob to Alice during the time interval $[t,t+1]$ as $f_{12}(t)$. In a similar way, Alice will also sell goods and services to Carol between $t$ and $t+1$ and will, as a result, earn an additional income of $f_{13}(t) \times x_3(t)$, where $f_{13}(t)$ is the fraction of $x_3(t)$ paid by Carol to Alice between $t$ and $t+1$ in exchange of the purchased goods and services. Thus, the total income earned by Alice over the time interval $[t,t+1]$ as a result of her doing business with other economic agents is:
\begin{equation}
    x_1(t+1) = f_{12}(t) \, x_2(t) + f_{13}(t) \, x_3(t) 
\end{equation}
\\
Let us now focus on Bob as a seller of some different type of goods and services to other economic agents, namely Alice and Carol who would then have to pay some fractions $f_{21}(t)$ and $f_{23}(t)$ of their respective incomes $x_1(t)$ and $x_3(t)$ at time $t$ to Bob. Thus, over the time interval $[t,t+1]$, Bob earns a total income of:

\begin{equation}
    x_2(t+1) = f_{21}(t) \, x_1(t) + f_{23}(t) \, x_3(t) 
\end{equation}
as result of him conducting business with the other agents.\\~\\

\noindent In a similar way, Carol, as a seller, would earn a total income of:  
\begin{equation}
    x_3(t+1) = f_{31}(t) \, x_1(t) + f_{32}(t) \, x_2(t) 
\end{equation}
over the time interval $[t,t+1]$, as a result of her selling other goods and services to Alice and Bob, $f_{31}(t)$ and $f_{32}(t)$ being the fractions of the buyers' respective incomes $x_1(t)$ and $x_2(t)$ that were paid to Carol. \\~\\
Using matrix algebra, the above three equations can be written as follows:
\begin{equation}
    \mathbf{x}(t+1) = \mathbf{F}_t \, \mathbf{x}(t) 
\end{equation}
where $\mathbf{x}(\tau) = [x_1(\tau), x_2(\tau), x_3(\tau)]^T$ is the income vector at time $\tau$, and the matrix:
\begin{equation}
    \mathbf{F}_{\tau} = [f_{ij}(\tau)]_{_{i,j\in \{1,2,3\}}} 
\end{equation}
is the $3 \times 3$ matrix whose entries are the above-mentioned fractions $f_{ij}(\tau)$ at time $\tau$. The superscript $\ ^T$ denotes, in the entire article, the transpose of a vector or a matrix. We should note that some of the entries in $\mathbf{F}_{\tau}$ might be equal to zero if there is no exchange between the corresponding agents. In particular, according to the above linear algebraic description of income circulation, the diagonal entries $f_{ii}(\tau)$ of the matrix $\mathbf{F}_{\tau}$ will be all zeroes, because agents don't sell things to themselves. But they can save money between one time instant and the next one. As business actors, they earn money by selling goods and services to others and, as living entities, they incur expenses to maintain themselves. The difference between the income earned and the expenses incurred would be the agents' savings. To be more specific, let us consider one individual agent $j$ which can be Alice, Bob or Carol. The total expenses that are incurred by $j$ between $t$ and $t+1$ can be calculated in the following way using the entries in the column $j$ of $\mathbf{F}_{\tau}$: 
\begin{equation}
    (\text{total expenses})_j = \sum_{\substack{i\in \{1,2,3\} \\ i \neq j}} f_{ij}(t) \, x_j(t) 
\end{equation}
Then, the difference:

\begin{equation}
\begin{split}
 x_j(t) - \, (\text{total expenses})_j & = x_j(t) \, - \sum_{\substack{i\in \{1,2,3\} \\ i \neq j}} f_{ij}(t) \, x_j(t)  \\
 & = \underbrace{\left( 1 - \sum_{\substack{i\in \{1,2,3\} \\ i \neq j}} f_{ij}(t) \right)}_{s_j(t)} \, x_j(t) \\
 \end{split}
\end{equation}
is what the agent $j$ would have saved over the time interval $[t,t+1]$.\\ 

\noindent Let us now redefine the diagonal entries $f_{jj}(\tau)$ for each agent $j \in \{1,2,3\}$ in the matrix $\mathbf{F}_{\tau}$ as: 

\[ f_{jj}(\tau) = s_j (\tau) = \left( 1 - \sum_{\substack{i\in \{1,2,3\} \\ i \neq j}} f_{ij}(\tau) \right)\]
and redefine the coordinates $x_i(\tau)$, $i\in \{1,2,3\}$, of the vector $\mathbf{x}(\tau)$ as the \textit{sum} of the income earned \textit{and} the savings made by agent $i$ between the time instants $\tau -1$ and $\tau$. With these new definitions of $f_{jj}(\tau)$ and $x_i(\tau)$, the matrix equation:  
\begin{equation}
    \mathbf{x}(t+1) = \mathbf{F}_t \, \mathbf{x}(t)  \label{Equ_8}
\end{equation}
remains valid with the understanding that the coordinates of $\mathbf{x}(\tau)$ at time $\tau$ represent the sum ``\textit{income} + \textit{savings}" made by the individual economic agents between $\tau-1$ and $\tau$, not just the \textit{income}. At this stage of building the linear algebraic model \eqref{Equ_8}, banks and financial institutions who would lend money to agents are not included. Because of that, the diagonal entries in the matrix $\mathbf{F}_{\tau}$ will remain non-negative. Note also that, with the new definitions of the diagonal entries and the coordinates of $\mathbf{x}$, the matrix $\mathbf{F}_{\tau}$ is column-stochastic.



\section{An agent-based income circulation model}

 The gedankenexperiment we used in the above discussion involved three agents \,---\, Alice, Bob and Carol. But we can easily repeat this experiment and generalize it to $n$ agents with $n>3$, in which case the final version of income circulation model becomes:\\

\begin{equation}
    \mathbf{x}(t+1) = \mathbf{F}_t \, \mathbf{x}(t)  \label{equ_9}
\end{equation}
where $\mathbf{F}_{\tau}$ is a square $n \times n$ non-negative matrix that is \textit{column-stochastic}, and that is referred to in this paper as the \textit{income circulation matrix} at time $\tau \geq 0$. It goes without saying that, for large values of $n$, the matrix $\mathbf{F}_{\tau}$ will likely have in it many entries that are zeroes, because it is very improbable in a large economy that every pair of agents engages in an act of selling and buying between the times $\tau$ and $\tau +1$.\\

\noindent Let us now specify the context within which the above matrix \eqref{equ_9} model would be valid:

\begin{enumerate}
    \item The economy at hand $\mathcal{E}$ is a closed economy made up of $n$ agents, and can be a small village, a region, or an entire country. The number $n$ can be in the hundreds, millions or billions of agents.
    \item An agent in this economy $\mathcal{E}$ can be:
    \begin{itemize}
        \item an individual who: (1) is an employee working for another agent (firm, individual, household, etc.), or (2) is running their own business (corner store, plumbing services, barber shop, handyman, and so on);
        \item a household made up of individuals who earn income, say the parents, and other ones who are dependent on the income earners;
        \item a firm which consists of a business or a collection of businesses that are, in principle, more sophisticated than the ones owned and operated by individuals. Those firms that are collections of businesses can be broken down into multiple agents in the matrix model we propose in this paper. For example, if the firm is a retail chain, then each retail store can be considered as a different agent in the income circulation matrix (ICM). Such a retail store sells goods and services to other agents in its neighborhood, while it buys goods and services from its corporate office which would be represented as a different agent in the ICM. In this respect, we should point out here that portions of the income circulation model (e.g., block matrices in the ICM that capture the interactions between the retail stores and their respective corporate offices) can be used effectively for the purposes of micro-economic analyses.    
    \end{itemize}
    
    \item At this initial stage of the development of the model, we assume that the agents in $\mathcal{E}$ cannot be banks or financial institutions who provide credit and, as a result, agent debt is not considered in this model. In general, the financial accounting concepts of \textit{liability} and \textit{asset} are intentionally omitted in the current model, but can be relatively easily considered in future versions of the model. Similarly, no governments collecting taxes and hiring civil servants are included at this stage in the current model. 
    
    \item Time is discretized into uniform steps $t$, $t+1$, $t+2$, etc. There is no restriction on the length of these time steps which can be in the range of seconds, hours, days, or weeks, depending on the intensity of the economic activity in $\mathcal{E}$.
    \item For each integer $i \in \{1, 2, \cdots, n\}$, let $x_i(\tau) \geq 0$ be the sum of income earned and savings made by agent $i$ between times $\tau - 1$ and $\tau$. We will refer to $x_i(\tau)$ as the \textit{wealth} of agent $i$ at time $\tau$. The vector $\mathbf{x}(\tau) = (x_1(\tau), x_2(\tau), \cdots, x_n(\tau))^T \in \Re_+^n$ is the wealth vector of agents in $\mathcal{E}$.
    
\end{enumerate}

\noindent Under the conditions specified in the above itemized list, the dynamics of the wealth vector  $\mathbf{x}$ are governed by the matrix equation \eqref{equ_9}. Using the rules of matrix algebra, we can show that the wealth vector $\mathbf{x}$ can be expressed as:
\begin{equation*}
    \mathbf{x}(t+1) = \left( \prod_{\tau=0}^{t} \mathbf{F}_{\tau} \right) \, \mathbf{x}(0)   
\end{equation*}
or, equivalently:\\
\begin{equation}
\boxed{\mathbf{x}(t) = \left( \prod_{\tau=0}^{t-1} \mathbf{F}_{\tau} \right) \, \mathbf{x}(0)}  \label{equa_10}
\end{equation}
where $\mathbf{x}(0)$ is the wealth vector of the economy $\mathcal{E}$ at an arbitrarily selected initial time instant $t=0$. The matrix equation \eqref{equa_10} constitutes what we believe to be a fundamental result about the dynamics of income circulation which, to our knowledge, has never been published in the economics literature, and which can be stated as follows: 
\vspace{0.0cm}
\begin{center}
\fbox{\begin{minipage}{25em}
\vspace{0.3cm}
\centering { \textit{The dynamics of wealth distribution in a closed economy where no credit is available to the economic agents are governed by \textbf{inhomogeneous products of column-stochastic matrices}.}}
\vspace{0.3cm}
\end{minipage}}
\end{center}
\vspace{0.2cm}

 The sum $M=\sum_{i=1}^n x_i(0)$ of the coordinates of $\mathbf{x}(0)$ is referred to in this paper as the \textit{monetary base}. Since the financial sector is  inexistent in $\mathcal{E}$, no money is created and, thus, $M$ remains constant over time. Using matrix algebraic rules, one can show that the sum of the coordinates of the wealth vector $\mathbf{x}(t)$ is invariant under transformations by inhomogeneous products of column-stochastic matrices, which ensures that the income circulation equation \eqref{equa_10} is consistent with the initial setup of $\mathcal{E}$. Also, in the cases where the length of the time steps are so short that no  business exchange takes place within a certain time interval $[\tau,\tau+1]$, the income circulation matrix $\mathbf{F}_{\tau}$ just collapses to the identity matrix, as agents get to keep all their wealth to themselves during this time interval. Because the identity matrix is the neutral element for matrix multiplication, the matrix model \eqref{equa_10} is consistent with times steps being of any size. We also conjecture that one can derive from the matrix model \eqref{equa_10} many of the empirical observations that economists have highlighted about the issues of economic inequality, including the fact that wealth  distributions in societies follow, under some conditions on the structures of the income circulation matrices $\mathbf{F}_{\tau}$ ($\tau >0$), the Pareto law which leads to the so-called 80-20 rule.\\

 Before we tackle the issue of the poor and vulnerable which is the main focus of this paper, the reader might be interested in the question of how the income circulation model \eqref{equa_10} can be scaled up to handle open economies. This question is discussed in Appendix B.
 
 \section{Societies can be cohesive or fragmented}
 Now that we have shown that the dynamics of wealth distribution in $\mathcal{E}$ follow patterns dictated by in-homogeneous products of stochastic matrices, one could choose to analyze the interactions between the society and vulnerable people using computer simulations and random matrix theory. While such approaches are highly valuable for future endeavors, they may not help us understand why and how things work the way they do. If computer simulations show that caring for the vulnerable is beneficial or not beneficial for the wealthy and for the society as a whole, we may not get much insights as to why and how that is the case. Because of this, we will keep the methods in this paper focused on deriving results from the ground up using deductive reasoning. To do that, we will assume in the remainder of this paper that the income circulation matrix $\mathbf{F}_{\tau}$ would keep changing over time, but only by small amounts, to the extent that we can consider that $\mathbf{F}_{\tau} \approx \mathbf{F}_{\tau +1}$ over a time interval $[0,\mathcal{T}] \ni \tau$ that is relatively long enough, in which case the income circulation model becomes: 

\begin{equation}
    \mathbf{x}(t) =  \mathbf{F}^{t} \, \mathbf{x}(0)  \label{equa_11}
\end{equation}
where $t \in [0,\mathcal{T}]$, $\mathbf{x}(0)$ is the wealth vector at the initial time $t=0$, and the matrix:
\[\mathbf{F}= \left[f_{ij} \right]_{\substack{i=1,n \\ j=1,n}}\] 
is a fixed column-stochastic matrix\footnote{\textbf{\ Note:} The assumption that $\mathbf{F}_{\tau}$ does not change by a lot over a certain time interval $[0,\mathcal{T}] \ni \tau$ is obviously a strong one. But, as it will be shown later in the paper, the analysis of the model $\mathbf{x}(t) =  \mathbf{F}^{t} \, \mathbf{x}(0)$ that results from this assumption is so insightful that it is worth it to consider the scenario where $\mathbf{F}_{\tau} \approx \mathbf{F}_{\tau +1}$ over the time interval $[0,\mathcal{T}]$. The theoretical analysis presented below can also be useful from a practical point of view; one way to make good use of the results derived in the subsequent sections from the approximation $\mathbf{F}_{\tau} \approx \mathbf{F}_{\tau +1}$ is to introduce the concept of \textit{average income circulation matrix}, $\mathbf{F}_{avg}$, over the time interval $[0,\mathcal{T}]$, and set $\mathbf{F}$ in the model \ref{equa_11} equal to $\mathbf{F}_{avg}$. Of course, the practical usefulness of using $\mathbf{F}_{avg}$ in the model \ref{equa_11} depends on how large the variations of $\mathbf{F}_{\tau}$ are over the interval $[0,\mathcal{T}]$. The authors believe that, in the case of large variations of $\mathbf{F}_{\tau}$, one could use the appropriate tools of differential calculus and formulas similar to Taylor series to come up with better approximations for $\mathbf{F}_{\tau}$ and take advantage of the results presented in the remainder of the paper. Finally, there is also the option of adjusting the size $\mathcal{T}$ of the $[0,\mathcal{T}]$ so that the approximations of $\mathbf{F}_{\tau}$ are good enough for practical purposes.}.
\\

 We then define the \textit{income circulation graph} of the economy $\mathcal{E}$ as the directed graph $\mathcal{G(\mathcal{E})}$ on the $n$ economic agents $\{1,2, \cdots, n \}$, in which there is a directed edge leading from agent $u$ to agent $v$ if and only if $f_{uv} \neq 0$. In other words, a directed edge exists from agent $u$ to agent $v$ when money flows out from agent $v$ to agent $u$ in exchange for some goods and services supplied by $u$ to $v$. A directed edge from agent $u$ to agent $v$ will be denoted as $u \rightarrow v$; $u$ is referred to as the initial agent of the edge, and $v$ the terminal agent. A \textit{path} in $\mathcal{G(\mathcal{E})}$ from agent $u$ to agent $v$ is a sequence of agents:\\

\[u = a_0, \, a_1, \, \cdots, \, a_{k-1}, \, a_k = v\]

\noindent such that $a_i \rightarrow a_{i+1}$ is a directed edge for each $i = 0, 1, \cdots, k-1$. The number $k$ of edges:
\[ a_0 \rightarrow a_1 ; a_1 \rightarrow a_2 ;  \dots  ; a_{k-1} \rightarrow a_k \]
in this path linking agent $u$ (\textit{initial} agent) to agent $v$ (\textit{terminal} agent) is called the \textit{length} of the path. In the remainder of the paper, the income circulation graph $\mathcal{E}$ may also be referred to as a \textit{business network}.  \\

\noindent We now introduce three types of societies based on the structure of the income circulation graph of the economy: a \textit{whole} society, a \textit{cohesive} society and a \textit{fragmented} society.  \\ 

 A society is said to form \textit{a whole} if the income circulation graph $\mathcal{G(\mathcal{E})}$ of the corresponding economy  $\mathcal{E}$ is strongly connected, which means that, for every pair $(u,v)$ of agents $u$ and $v$ in the graph $\mathcal{G(\mathcal{E})}$, there exists a path $\mathcal{P}(u \rightarrow v)$ leading from agent $u$ to agent $v$. The length $k$ of this path linking $u$ to $v$ may vary with the pair $(u,v)$. For those pairs of agents who work closely with each other, this length could be small, while for other pairs, it might be larger. A question that arises here is this:
\begin{center}
\textit{Would it be possible, in a society that forms a whole, to connect \\
any agent $u$ to any other agent $v$ with a path \\of a pre-determined length $k_0$ that is independent of the pair $(u,v)$?} 
\\
\end{center}

\noindent Stated in a more concrete fashion, this question can be rephrased as follows:
\begin{center}
\textit{Whether agents $u$ and $v$ are both billionaires \\ in the top layers of the society, \\ \textit{or} \\ $u$ is billionaire in the top social class and $v$ is a parking lot attendant \\ at the bottom of the social hierarchy, \\ is it still possible to connect $u$ and $v$ with a path of \\ the \underline{same} length in the income circulation graph of the economy?} \\~\\
\end{center}

 \noindent The answer depends on the structure of the income circulation graph $\mathcal{G(\mathcal{E})}$, which is of course dictated by the structure of the income circulation matrix $\mathbf{F}$. Consider for example a simple economy $\mathcal{E}_{ex}$ of 3 agents where the income circulation matrix is as follows:
\[ \mathbf{F}_{ex} =
\begin{bmatrix}
0 & 1 & 0\\
0 & 0 & 1\\
1 & 0 & 0
\end{bmatrix} \]

\noindent Agent 3 buys everything they need from agent 2, agent 2 buys everything they need from agent 1, and agent 1 buys everything they need from agent 3. The corresponding income circulation graph is as follows:
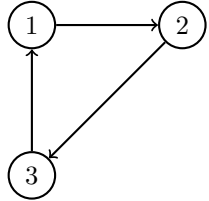
\vspace{1cm}

\begin{center}
    
\begin{tikzpicture}[node distance={20mm}, thick, main/.style = {draw, circle}] 
\node[main] (1) {$1$};
\node[main] (2) [right of=1] {2};
\node[main] (3) [below  of=1] {3}; 
\draw[->] (1) -- (2);
\draw[->] (2) -- (3);
\draw[->] (3) -- (1);
\end{tikzpicture}
\end{center}

\vspace{1cm}
\noindent The society that underlies this economy does form a \textit{whole} because, for every $(i,j)\in \{1,2, 3\} \times \{1,2, 3\}$, we can find a path that leads from $i$ to $j$ in the graph. Table  \ref{tab:1} in Appendix C lists these paths, along with their respective lengths.  
\\

 However, based on the information provided in this table, it does not seem possible to connect pairs of agents using paths of \textit{\textbf{one}} same length, even if we attempt to increase the number of edges in these paths. A quick inspection of the third column in table \ref{tab:1} shows that the lengths of the paths that connect an agent to itself exhibit a periodicity of 3, which is exactly the same as the number of eigenvalues of the  matrix $\mathbf{F}_{ex}$ on the unit circle. But this is no accident. There is an entire class of stochastic matrices, called periodic or imprimitive matrices, that exhibit periodic features and that have been studied extensively in the scientific literature \cite{debreu, seneta,gantma}. We conjecture that, when the power $\mathbf{F}^{t}$ in the model \eqref{equa_11}\footnote{or, for that matter, the in-homogeneous product $\left( \prod_{\tau=0}^{t-1} \mathbf{F}_{\tau} \right)$ in the general model \eqref{equa_10}} is and remains mostly imprimitive over a period of time $[0,\mathcal{T}] \ni t$ that is long enough, the income  circulation models proposed in this paper\footnote{in conjunction with robust scientifically-based frameworks that help identify and analyze cause-and-effect relationships.} can provide first-principle explanations for the so many cyclical phenomena we often observe in the real-world, including economic cycles, cyclical unemployment, innovation cycles and so on.\\ 


\noindent In the other situation where the matrix $\mathbf{F}$ in the income circulation model \eqref{equa_11} is an aperiodic matrix, also known as a primitive matrix, the following property \textbf{P} will hold true:\\

\hspace{1cm} \begin{minipage}{9.6cm}
\begin{center}
\begingroup \textit{    
there exists an integer $k_0$ such that any agent $u$ in the income circulation graph $\mathcal{G(\mathcal{E})}$ can be reached from any other agent $v$ in this same graph by traversing $k_0$ edges in the direction in which they point, regardless of where these agents are situated in the social class hierarchy.}
\endgroup
\end{center}
\vspace{-0.4cm}
\begin{center}
    (Property \textbf{P})
\end{center}

\end{minipage}
\\~\\

 The integer $k_0$ is not dependent on any particular agent or pair of agents in the economy. Rather, it is a characteristic of the economy $\mathcal{E}$ as whole, and the structure of its income circulation graph $\mathcal{G(\mathcal{E})}$. In this paper, we will refer to a society where the income circulation graph $\mathcal{G(\mathcal{E})}$ of its underlying economy $\mathcal{E}$ satisfies the property \textbf{P} as a \textit{cohesive whole}, or simply \textit{cohesive}. The smallest possible value of the integer $k_0$ referred to in the property \textbf{P} is called, in the literature on nonnegative matrices, as the \textit{exponent} of the primitive matrix $\mathbf{F}$. In this paper, we will refer to it as the \textit{number of degrees of separation} in the business network represented by the graph $\mathcal{G(\mathcal{E})}$, and use it to assess how cohesive a society is. Intuitively, the smaller the number of degrees of separation $k_0$, the closer the agents are to each other in the graph $\mathcal{G(\mathcal{E})}$, and the more cohesive the society is. The adjectives ``\textit{closer}'' and ``\textit{cohesive}'' are to be understood here from a \textit{trading} and \textit{business} perspective, not ethnic, religious or cultural perspectives. The questions of how \textit{business cohesiveness} affects or gets affected by \textit{cultural cohesiveness} are of course valid and important ones, but they are outside the scope of this paper which focuses \textit{solely} on the former type of cohesiveness. \\

 Because the magnitude of $k_0$ and the level of business cohesiveness of the society change in opposite directions, we introduce in what follows the number:
\begin{equation}
    \mathcal{C} = 1/k_0 \in ]0,1]   \label{equa_14}    
\end{equation} 
as a measure of the \textit{business cohesiveness} of the society. The \textit{larger} the value of $\mathcal{C}$, the smaller the number of degrees of separation in the graph $\mathcal{G(\mathcal{E})}$, the closer the agents are to each other, and the \textit{higher} the level of business cohesiveness of the  society. In the exceptional situation where every agent trades \textit{directly} with every other agent in the society, the society will reach the highest possible cohesiveness, and $\mathcal{C}$ will hit its maximum value which is 1, because every entry $f_{uv}$ in the matrix  $\mathbf{F}$ will then be nonzero and therefore the length of the paths connecting agents to each other will be equal to 1. This situation may happen in a village of a relatively small number of agents, because these agents will be geographically close to each other. But in a large country, it will be difficult to realize this situation where agents all trade directly with each other. This will be even more difficult in the global economy because of the geographic constraints, although digital commerce and platforms have nowadays allowed agents from around the world to defy these constraints and trade directly with each other. It is important, however, to point out that geographic constraints are \textit{not} the only obstacle in achieving a greater business cohesiveness. As it will be highlighted in the sections below, the level of trust among agents in a society would play a major role in the process of building cohesiveness.\\

 In the remainder of the paper, we will refer to a society that does not form a whole as a \textit{fragmented} society. One intriguing question that the reader may wonder about at this stage is the following: \textit{what will it take for a society that forms a whole to become a cohesive whole}? It takes only one thing: \textit{just one agent $u$ in the economy is able to save money for themselves, i.e., $f_{uu} \neq 0$}. Various mathematical proofs of such a statement are available in the literature on nonnegative matrices (e.g. \cite{solow}, page 40). Note that this condition, $\exists u \in \{1, 2, \cdots, n\}$ such that $f_{uu} \neq 0$, is sufficient for a whole society to be cohesive, but not necessary. \\ 

 The other question that we like to address before we move to the study of the vulnerable in society is about the order of magnitude of $k_0$. The number of degrees of separation in social networks that use chains of `friend of a friend' statements to connect people has been reported to be low and popularized as `six-degrees of separation' \cite{watts}, but \textit{what would this number of degrees of separation be in business networks that use the actual act of selling and buying, not just friendship, to connect people}? Building trading relationships that involve an exchange of money is certainly more demanding and requires more commitment than friendship relationships. Because of that, one would expect that \textit{the degrees of business separation} $k_0$ to be of a higher order of magnitude than \textit{the degrees of social separation}. Mathematically, we can show that $k_0$ for a cohesive society of $n$ agents would, in the worst case, be $(n-1)^2 +1$ \cite{wielandt}. Hence, for an economy of, say, 1 million agents, the value of $k_0$ would be, at most, of the order of 1 trillion degrees. But, if the economy has $\nu>0$ agents who are able to save money for themselves, i.e., the income circulation matrix has $\nu>0$ diagonal entries $f_{ii}$ that are non-zero, then $k_0$ would be at most $2n - \nu -1$ (see for example \cite{dulmage}), which means that $k_0$ becomes of the order of $n$, not $n^2$ anymore. If every single agent in the economy is able to save money for themselves, then $k_0$ goes down to $n-1$ in the worst case scenario. These results are directly relevant for the final conclusion we prove later on in this paper, because the rate at which the top wealthy class recovers its support money goes up significantly when the degrees of business separation $k_0$ in a cohesive society decreases.

 \section{Wealthy agents in cohesive societies benefit from supporting the vulnerable}

Here we show that, in cohesive societies, it is beneficial for wealthy agents to reach out to vulnerable people in the society and support them. In other words, those acts that we often label as altruistic or selfless and that consist in helping the poor and vulnerable in society should actually be viewed by wealthy individuals as selfish acts, as long  as the society is a cohesive whole. The expression ``\textit{support for the poor and vulnerable}'' refers here to the act of letting a small amount of money flow out of the top social layers made up of wealthy individuals to the very bottom layers of the society, \textit{without} obtaining any goods or services in exchange for this money.\\  

 If the society is fragmented, however, it will be difficult to conclude that wealthy agents would benefit from supporting the poor and vulnerable, although they may choose to provide such a support for religious and purely moral reasons. Stated differently, wealthy individuals who wish to support those at the very bottom layers of the society would be more effective if they first focus on improving the cohesiveness of the society.  \\

 To start our reasoning, let us re-order the rows and columns of the income circulation matrix $\mathbf{F}$ in such a way that the wealthy agents appear towards the top of the matrix (i.e., wealthy agents are labeled by the first values of the subscript $i \in \{1,2 ,3 \cdots, n\}$), while the ones who are extremely poor and vulnerable are at the bottom of the matrix (i.e., poor people are labeled by the last values of the subscript $i$). With this re-organization of the matrix, the rows  at the very top of the matrix will have the largest (row) sums and/or the largest numbers of non-zero entries. Note, however, that row sums and numbers of non-zero  entries in a row are not an \textit{exact} indication of which agent will be wealthier and which one will be less wealthy.\\

\noindent To analyze the interactions between the society and the vulnerable, let us also structure $\mathbf{F}$ as a block matrix in the following way:\\

\begin{equation}
\renewcommand{\arraystretch}{1.5} 
\mathbf{F}=  \left[\begin{array}{@{}c|c@{}}
  \mathbf{F}_{1,1} & \mathbf{b}
\\ \hline
\mathbf{c}^T 
  & d
\end{array}\right] = 
\left[\begin{array}{@{}c|c@{}}
  \mathbf{F}_{1,1} &
  \begin{matrix}
  b_1 \\
  \vdots \\
  b_{n-1}
  \end{matrix}
\\ \hline
  \begin{matrix}
  c_1 \quad \cdots \quad c_{n-1}
  \end{matrix}
  & d
\end{array}\right]   \label{equa_15}
\end{equation}
where

\begin{itemize}
    \item $\mathbf{F}_{1,1}$ is the $(n-1) \times (n-1)$ matrix of income circulation coefficients $f_{ij}$ for the agents $1, 2, \cdots, n-1$
    \item the coordinates of the the vector $\mathbf{b} = [b_1 \quad \cdots \quad b_{n-1}]^T$ represent the expenditures that agent $n$ pays for goods and services purchased from the other $n-1$ agents. More specifically, for each $i \in \{1, 2, \cdots, n-1\}$, the coordinate $b_i$ is the fraction of agent $n$'s wealth spent on buying things from agent $i$. 
    \item the coordinates of the vector $\mathbf{c} = [c_1 \quad \cdots \quad c_{n-1}]^T$ represent sources of income that agent $n$ earns by selling his goods and services to the other $n-1$ agents in the society. For each $j \in \{1, 2, \cdots, n-1\}$, the coordinate $c_j$ is the fraction of agent $j$'s wealth paid to agent $n$ in exchange for some goods and services sold by this agent $n$.
    \item  $d \in [0,1]$ represents the income fraction that agent $n$ saves for herself from one time step to the next.
\end{itemize}

 In the discussions that follow, we will assume that agent $n$ at the very bottom of the matrix is a  typical case of an extremely poor and/or highly vulnerable person in the society. As a result, most of  the entries in vector $\mathbf{c}$ will be zero, and those entries that are non-zero in this vector will probably have very small values. Let us consider the  entry $c_1$ for example. If agent 1 decides to hire agent $n$, and agent 1 is among the wealthiest in society (as a result of the re-ordering discussed above for matrix $\mathbf{F}$'s columns and rows), the market conditions are probably such that agent 1 is more likely to pay something closer to $c_1= 0.000001\%$ rather than $c_1=0.1\%$ of his or her wealth to agent $n$ in exchange of, say, nanny or cleaning services. On the other hand, the entries in the vector $\mathbf{b}$ may probably have higher values or, if they are not so high, the sum $\sum_{i=1}^{n-1}  b_i$ could probably be close to 1, in which case $d$ may be almost 0 (i.e., agent $n$ is so poor that he is unable  to save money from one time  period to the next). This is because the entries $b_i$ represent fractions of agent $n$'s wealth (which is very small) spent on buying food, shelter, and other necessities from the agents $j\in \{1, 2, \cdots, n-1 \}$ in the economy to cover the needs of this poor agent $n$. \\

 Under some abnormal behavior conditions, however, the entries in the vector $\mathbf{b}$ could also be all zeroes, in  which case $d$ will be equal to 1. This would mean that agent $n$ buys no goods or services from  the economy he belongs to ($b_i= 0$ for all $i \in \{1, 2, \cdots, n-1\}$), although he is earning income from  this very  economy ($c_i \neq 0$ for at least one $i \in \{1, 2, \cdots, n-1\}$). Of course, such a situation goes against the physical reality, which stipulates that agent $n$, being a human being, would require food, shelter, and other basic necessities to stay alive. But it is  possible that this agent $n$ resorts to begging and other less dignifying acts (using other people's leftovers, shoplifting, etc.) to satisfy these necessities, while saving his cash in its entirety  (i.e., $d=1$). In  what follows, we will refer to such an individual, for whom $\mathbf{b} = \mathbf{0}$ but $\mathbf{c} \neq 0$, as a \textit{pure cash hoarder}. \\

 Supporting a cash hoarder, even if he is a poor and vulnerable, is not beneficial to wealthy agents, nor is it to the whole economy. As shown in the theorem \ref{theorem_1}  below, if there is a cash hoarder in a closed economy $\mathcal{E}$, then,  over the long-term, this cash hoarder will end up absorbing all of the cash in the economy. 


\newtheorem{thm}{Theorem}[subsection]
\renewcommand{\thethm}{\arabic{thm}}

\begin{theorem} 
With the notations in equation \eqref{equa_15}, assume that:
\begin{itemize}
    \item agent $n$ is a pure cash hoarder, i.e.,  $\mathbf{b} = \mathbf{0}_{n-1,1}$ but $\mathbf{c} \neq \mathbf{0}_{n-1,1}$. (\textit{Note that because $\mathbf{b} = \mathbf{0}_{n-1,1}$, the value of $d$ is 1}).   
    \item the rest of the economy run by other agents $\{1, 2, \cdots, n-1\}$ is a whole, i.e., the income  circulation graph corresponding to the matrix  $\mathbf{F}_{1,1}$ is strongly connected.    
\end{itemize}
Then, for all $k \geq 1$: 

\begin{equation}
\renewcommand{\arraystretch}{1.5} 
\mathbf{F}^k=
\left[\begin{array}{@{}c|c@{}}
  \mathbf{F}_{1,1}^k & \mathbf{0}_{n-1,1}
\\ \hline
\mathbf{c}^T \,  \sum_{i=0}^{k-1} \mathbf{F}_{1,1}^i  
  & 1
\end{array}\right]
\end{equation}
Furthermore:
\begin{equation}
    \lim_{k \to +\infty}  \mathbf{F}^k  = \left[\begin{array}{@{}c|c@{}}
  \mathbf{0}_{n-1,n-1} & \mathbf{0}_{n-1,1}
\\ \hline
\mathbf{c}^T \, (I-\mathbf{F}_{1,1})^{-1}  
  & 1
\end{array}\right]
\end{equation}
\label{theorem_1}
\end{theorem}
\begin{proof}
See Appendix D.
\end{proof}

 In the real-world, however, the society will of course start identifying problems way before the cash ends up in the hands of a few individuals. In advanced economies such as the United States and Canada, every now and then, individuals are caught collecting food stamps and other benefits, while owning  millions of dollars in the  bank. In other countries such as India for example, demonetisation initiatives get implemented to punish those individuals who hoard (ill-gotten) cash.   \\

 Love of  money could be one reason why people become cash hoarders. They perceive money as the solution to all problems and tend to save very large portions of it, even if it is at the expense of their own well-being and the cohesion of their society. Cash hoarding behavior could also be a result of the distrust which tends to build up in societies that reject some of its citizens and offer them no opportunities. In the income circulation matrix, such a rejection manifests itself for agent $n$ when the vector $\mathbf{c}$ has all of its entries equal to zero. There are many countries around the world where marginalized people earn no income from the economy within which they live, and end up losing faith in their respective countries. Some of these  marginalized people end up joining groups of migrants, such as the ones who attempt to cross the Mediterranean sea from Africa to Europe and often lose their lives in their migration journeys. Other ones may stay in their country and, once they have an opportunity to earn income ($c_i \neq 0$ for some $i \in  \{1, 2, \cdots, n-1\}$), they would resort to cash hoarding, as a way to manage future risks. When there is a large number, say $m$, of such marginalized people in the society, the income circulation matrix $\mathbf{F}$ takes the following form: 

\begin{equation}
\renewcommand{\arraystretch}{1.5} 
\mathbf{F}=  \left[\begin{array}{@{}c|c@{}}
  \mathbf{F}_{1,1} & \mathbf{F}_{1,2}
\\ \hline
\mathbf{F}_{2,1} 
  & \mathbf{F}_{2,2}
\end{array}\right]  \label{equa_16}
\end{equation}
where:
\begin{itemize}
    \item $\mathbf{F}_{1,1}$ is the $(n-m) \times (n-m)$ income circulation matrix of the $n-m$ wealthiest agents (individuals, households or businesses) in the economy;
    \item  $\mathbf{F}_{2,2}$ is the $m \times m$ income circulation matrix of the $m$ marginalized people;
    \item $\mathbf{F}_{2,1}$ and $\mathbf{F}_{1,2}^T$ are $m \times (n-m)$ matrices whose entries are (almost) all zeroes, because the wealthy agents $\{1, 2, \cdots, n-m\}$ hardly buy any goods or services from the marginalized agents $\{n-m+1, \cdots, n\}$, and the marginalized agents $\{n-m+1, \cdots, n\}$ either hoard the little cash they can  earn or trade only among themselves. 
\end{itemize}

 A society whose economy is governed by such an income circulation matrix is obviously not cohesive, and there are no physics-based reasons for wealthy individuals to reach out to and support the poor and vulnerable in the lower tiers $\{n-m+1, \cdots, n\}$ of this society. In fact, it could be risky for wealthy individuals in such a society to allow a steady flow of their cash to be transferred to the poor agents to help support them, because of the distrust that tends to build up over time in a fragmented society. Here are indeed some asymptotic results that one can prove mathematically, using a similar approach as the one outlined in Appendix D:
\begin{enumerate}

    \item if $\mathbf{F}_{2,1} \neq \mathbf{0}_{m,n-m}$ and $\mathbf{F}_{1,2} = \mathbf{0}_{n-m,m}$  (which means that the wealthy agents are engaging the poor ones in jobs, but the poor agents are distrustful of the wealthy agents, and are thus hoarding their cash or trading only among themselves), then the poor agents will asymptotically end up sucking all of the cash in the economy; 
    \item if $\mathbf{F}_{1,2} \neq \mathbf{0}_{n-m,m}$ and  $\mathbf{F}_{2,1} = \mathbf{0}_{m,n-m}$ (which means that the poor  agents are buying goods and services from the wealthy ones, but the wealthy agents are distrustful of the poor ones, and are thus not hiring them at all), then the wealthy agents will asymptotically end up sucking all of the cash in the economy.
\end{enumerate}
Asymptotic analyses are useful, as they provide us with insights on how the system would keep evolving if its structure remains the same over time. They also help us understand the nature, direction and driving forces of the dynamics. In real-world situations, however, human beings are able to detect, using various means, the direction of the changes that are affecting their socio-economic situations and react to them. The cases of matrices $\mathbf{F}_{2,1}$ and $\mathbf{F}_{1,2}$ becoming intermittently equal to zero, as a result of class struggle, highlight the type of class struggle mechanics that could arise in fragmented societies. In some instances, consumer boycotts that are aimed at big businesses in the economy could be a manifestation of such class struggle mechanics.\\ 

 Getting a society out of a fragmented state into a cohesive one where both matrices $\mathbf{F}_{2,1}$ and $\mathbf{F}_{1,2}$ become nonzero and stay that way in a sustainable fashion is a challenging game-theoretic problem outside the scope of this research. In the rest of the paper, we will assume that the society is a cohesive whole, and thus none of the above situations  would arise. Below, we will show that, in a cohesive society, wealthy agents at the top of the social class hierarchy will only gain by letting tiny small fractions of their wealth flow down all the way to the very bottom of the social class hierarchy for the following two reasons:

\begin{enumerate}
    \item Below we will prove mathematically that the small amount of money that wealthy agents may let flow to the very bottom tiers of the society, to help support the poor and vulnerable, will eventually go back up to the top layers of the social hierarchy. Thus, from the monetary point of view, the high class of wealthy agents does not lose, when it reaches out to the poor and vulnerable and support them.      
    \item From a business point of view, wealthy agents will gain by helping the very poor and vulnerable, because such a help will contribute to and reinforce the trust within the society, promote cohesion among the social layers, and engage poor people as customers for the businesses in the economy which would be owned mostly by the wealthy agents.  
\end{enumerate}
 In what follows, we will focus on the ideal case where the income circulation matrix $\mathbf{F}$ remains constant over time, but extensive computer simulations, where Gaussian perturbations of the entries $f_{i,j}$ are introduced at each time step, showed that the result proven below remains valid for in-homogeneous products of stochastic matrices $\prod_{\tau=0}^{t} \mathbf{F}_{\tau}$, as long as these products of matrices remain primitive (i.e., the society remains a cohesive whole). Focusing on the ideal case of a constant income circulation matrix $\mathbf{F}$  helps understand the mechanisms of  why and how supporting the vulnerable in a cohesive society is beneficial to the wealthy agents, to the poor ones and to the society as a system. \\

\noindent As discussed in the above sections, we will assume that the rows and columns of the income circulation matrix $\mathbf{F}$ have been ordered in such a way that:

\begin{enumerate}
    \item the top rows $i \in H=\{1, 2, \cdots, h\}$ of the matrix correspond to the very wealthy agents in the high classes of the society. The number $h$ would be much smaller than $n$. For example, $h$ might be in the range of $10\%$ or $15\%$ of $n$, or perhaps lower than that, depending on how concentrated the wealth is at the top of the society.
    
    \item the very bottom rows $i \in L=\{l, l+1, \cdots, n\}$ correspond to the extremely poor individuals in the lower tiers of the society who might be in abject poverty. Depending on the overall health of the economy and the proportion of agents in the middle class, one could think of the number $l$ to be in the range of $0.8\, n$ or $0.9 \, n$. 
\end{enumerate}

 Let us now assume that, at a certain time instant $t_0$, a wealthy agent $h_0 \in H$ from the very top class of the society removes a small amount $\epsilon$ from  her wealth $x_{_{h_0}}(t_0)$, and passes it along to one of the poorest agents $l_0 \in L$ whose wealth at that time instant $t_0$ becomes $x_{_{l_0}}(t_0) + \epsilon$. In what follows, we will refer to this action of removing $\epsilon$ from a wealthy agent's wealth and transmit it to one of the poorest agents as the $\epsilon$-\textit{support method}. When the $\epsilon$-support method is applied between agents $h_0 \in H$ and $l_0 \in L$ at time $t_0$, the wealth vector in the economy $\mathcal{E}$ at this time $t_0$ becomes:

\begin{eqnarray}
    \mathbf{x}_{\epsilon}(t_0) = & [x_{_1}(t_0) \ \textbf{,} & \cdots \ \textbf{,} \ x_{_{h_0}}(t_0) - \epsilon \ \textbf{,} \  x_{_{(h_0+1)}}(t_0) \ \textbf{,} \  \nonumber \\
    &  & \cdots \ \textbf{,} \ x_{_{l_0}}(t_0) + \epsilon \ \textbf{,} \ x_{_{(l_0+1)}}(t_0) \ \textbf{,} \ \cdots \textbf{,} \ x_{_n}(t_0)]^T \label{equa_18}
\end{eqnarray}

\noindent What will happen to the income circulation process in the society after the $\epsilon$-support method is applied at time $t_0$? Before we address this question mathematically, let us first look at it in an intuitive manner. \\ 

 When the poor agent $l_0$ at the bottom of the social hierarchy receives an additional amount $\epsilon$  at $t_0$, she will end up spending it on  basic necessities of life because: 

\begin{enumerate}
    \item as discussed above, her income is very low (very small row entries, i.e., either $f_{_{l_0,j}} = 0$ or $f_{_{l_0,j}} \ll 1$) and can hardly meet her expenditures on basic needs (some of the column entries are $f_{_{i,l_0}}$ are high and/or the sum $\sum\limits_{\substack{i=1,n\\ i\neq l_0}}  f_{_{i,l_0}}$ is close to 1)
    \item the society is cohesive with no distrust or mistrust among its agents and, as a result, agent $l_0$ will not resort to cash hoarding and acts such as begging or shoplifting.     
\end{enumerate}

 If agent $l_0$ spends the amount $\epsilon$, say, at the local grocery store to buy food items and other basic necessities, then this local store will end up buying proportionally more goods from, say, a wholesaler to replenish his store's shelves; the wholesaler will then order more goods from manufacturing businesses, who will then buy more tools and services such as insurance policies and marketing campaigns, and so on. These manufacturing businesses, wholesalers and service  providers would all stack up at higher levels (than agent $l_0$'s) in the social hierarchy, and their corresponding rows in the income circulation matrix $\mathbf{F}$ of the economy are higher up in this matrix. As a result, the amount $\epsilon$ that was passed on to agent $l_0$ to support him  will \textit{end up rising back up to the higher tiers of the social hierarchy}, and will \textit{eventually make it into the coffers of those wealthy agents} who have a larger number of nonzero entries in their respective rows in the matrix $\mathbf{F}$. It should be highlighted, however, that, when the amount $\epsilon$ rises back up to the higher social tiers, it will \textit{not} necessarily end up in the coffers of the wealthy agent $h_0$ that had initially removed it from her wealth $x_{_{h_0}}(t_0)$ at time $t_0$, and used it to support agent $l_0$. In this respect, the dynamics of money circulation will have more of a statistical nature than a deterministic one.   \\

\noindent Quantitatively, the dynamics of the wealth vector $\mathbf{x}_{\epsilon}(t_0 +k)$, with integer $k > 0$, after the time $t_0$ when the $\epsilon$-support method was applied between agents $h_0$ and $l_0$, is governed by the following matrix equation:
\begin{equation}
    \mathbf{x}_{\epsilon}(t_0 +k) =  \mathbf{F}^k \, \mathbf{x}_{\epsilon}(t_0)   \label{equa_19}
\end{equation}

 An in-depth discussion of the meaning of \textit{small} in the expression ``small amount $\epsilon$'' is of course required. For now, however, we will assume that $\epsilon$ is small enough to the extent that the changes that may take place in the structure or entries of the matrix $\mathbf{F}$ at times $t_0 + k$ ($k>0$), as a result of the new actions of agents $h_0$ and $l_0$ (who just exchanged money in an unusual way at $t_0$), remain \textit{negligible}. For example, situations such as the 75-percent super-tax imposed by the  Government of France on the rich in 2012 does clearly \textit{not} fit in this context, because taking 75\% of the wealth of a rich individual and giving it away to poor individuals or spending it on other items will certainly change the behaviors and actions of the agents who take part in this transaction, including the rich and poor individuals that are involved. Hence, in what follows, we consider that $\epsilon \ll x_{_{l_0}}(t_0) \ll x_{_{h_0}}(t_0)$ so that $\mathbf{F}$ does not change much after the $\epsilon$-support method is applied between agents $h_0 \in H$ and $l_0 \in L$ at time $t_0$.\\

\noindent The wealth vector $\mathbf{x}_{\epsilon}(t_0)$ from equation \eqref{equa_18} can be re-expressed as follows: 
\begin{eqnarray}
    \mathbf{x}_{\epsilon}(t_0) = \mathbf{x}(t_0) - \epsilon \,  \mathbf{e}_{_{h_0}} + \epsilon \,  \mathbf{e}_{_{l_0}} 
\end{eqnarray}
where $(\mathbf{e}_1, \mathbf{e}_2, \cdots, \mathbf{e}_n)$ represent the standard basis of the $n$-dimensional space spanned by wealth vectors. Thus, the equation \eqref{equa_19} can be re-written as follows:
\begin{equation}
    \mathbf{x}_{\epsilon}(t_0 +k)  =  \mathbf{F}^k \,\mathbf{x} (t_0) + \mathbf{F}^k \, \left(- \epsilon \,  \mathbf{e}_{_{h_0}} + \epsilon \,  \mathbf{e}_{_{l_0}} \right) = \mathbf{x}(t_0 +k) + \epsilon \, 
\mathbf{F}^k \, \left(\mathbf{e}_{_{l_0}} - \mathbf{e}_{_{h_0}} \right) \label{equa_22}   
\end{equation}

\noindent In the following paragraphs, we analyze the second term $\epsilon \, 
\mathbf{F}^k \, \left(\mathbf{e}_n - \mathbf{e}_1 \right)$ of the above equation.  \\

\noindent The first observation we make here is that the income circulation matrix $\mathbf{F}$, which should be viewed as a linear operator on the $n$-dimensional space spanned by wealth vectors, conserves the sum of the coordinates of any vector $\mathbf{x}  =  [x_1, x_2, \cdots, x_n]^T \in \Re^n$: 
\begin{equation}
    \sum_{i=1}^n (\mathbf{x})_i  = \sum_{i=1}^n (\mathbf{F}\, \mathbf{x})_i
\end{equation} 
This is because the sum of each of the columns of $\mathbf{F}$ is equal to 1.  As a result, the length of any \textit{non-negative} vector, measured using the $l_1$ norm, will also be conserved, since its $l_1$  norm  is  equal  to the sum of its (non-negative) coordinates. But, when $\mathbf{F}$ is applied repeatedly to a vector such as  $(\mathbf{e}_{_{l_0}} - \mathbf{e}_{_{h_0}})$, the length of this vector will keep shrinking again and again, until it becomes almost 0. We will show herein that the dwindling of the length of $\mathbf{F}^k \, (\mathbf{e}_{_{l_0}} - \mathbf{e}_{_{h_0}})$ as $k$ keeps increasing is a direct consequence of the business cohesiveness of the society that underpins the economy $\mathcal{E}$, and that the stronger this cohesiveness is, the faster the length of $\mathbf{F}^k \, (\mathbf{e}_{_{l_0}} - \mathbf{e}_{_{h_0}})$ will vanish.

\section{How business cohesiveness leads to the dwindling of the length of $\mathbf{F}^k \, (\mathbf{e}_{_{l_0}} - \mathbf{e}_{_{h_0}})$ as $k$  increases}

 Since the society at hand is a cohesive whole, there exists an integer $k_0$ such that any agent $u$ can be reached from any other agent $v$ by traversing $k_0$ edges in the direction in which they point in the income circulation graph ($k_0$ being the number of degrees of business separation, also known as the exponent of the matrix $\mathbf{F}$). Based on the rules of matrix multiplication in relation to the properties of graphs (see for example \cite{ brualdi}, theorem 3.1.2, page 51), this characteristic of a cohesive society implies that all the entries in the matrix $\mathbf{F}$ to the $k_0\mbox{-}$th power will be positive.  That is:
\[ \forall (i,j) \in \{1, \cdots, n\} \times \{1, \cdots, n\},  (\mathbf{F}^{k_0})_{i,j}  >  0\]

\noindent Let us define $\mathbf{G} = \mathbf{F}^{k_0}$, and $g_{ij}$ where $(i,j) \in \{1, \cdots, n\} \times \{1, \cdots, n\}$ as the entries in the matrix $\mathbf{G}$. Let us also define the number $\alpha_i>0$, for each $i\in \{1, \cdots, n\}$, as the minimum of the entries in row $i$ of $\mathbf{G}$:
\[ \alpha_i = \min_{j=1,n} g_{_{i,j}}  \] 
This number $\alpha_i \in \, ]0,1[$ can be thought of as an indication of the smallest fraction of someone's wealth that the economy $\mathcal{E}$ is willing to pay to agent $i$ for the goods and the services he is selling. We will refer to it as the \textit{generosity coefficient} of the economy towards agent $i$. The sum $g = \sum_{i=1}^n \alpha_i \in \, ]0,1[$ of the generosity coefficients of all agents in the economy is referred to as the \textit{overall generosity level} of the economy, and the matrix $\mathbf{G}$ as the \textit{generosity matrix} of the economy. Using proper techniques for finding tight bounds, we can show that:  

\begin{equation}
    \norm{\mathbf{G}  \, \mathbf{u}} \leq (1 - g) \, \norm{\mathbf{u}} \label{equa_24}
\end{equation}
for any vector $\mathbf{u} = [u_1, u_2, \cdots, u_n]^T$ that satisfies $\sum_{i=1}^n u_i = 0$. The lengths $\norm{\mathbf{G}  \, \mathbf{u}}$ and $\norm{\mathbf{u}}$ are measured using the $l_1$ norm (See the proof in the Appendix E).\\

\noindent Let us now look at the behavior of the sequence of vectors $\mathbf{F}^k \, (\mathbf{e}_{_{l_0}} - \mathbf{e}_{_{h_0}})$ as $k$ increases. Using Euclidean division concepts, the number $k$ can be expressed as: $k= q \, k_0 + r$, where $q$ is an integer that increases with $k$, and $0\leq r  < k_0$.  Then:

\begin{equation}
\begin{split}
    \mathbf{F}^k \, (\mathbf{e}_{_{l_0}} - \mathbf{e}_{_{h_0}})  & =  \mathbf{F}^{q\,k_0 + r} \, (\mathbf{e}_{_{l_0}} - \mathbf{e}_{_{h_0}})\\
     &  =  \left(\mathbf{F}^{k_0}\right)^q \mathbf{F}^r \, (\mathbf{e}_{_{l_0}} - \mathbf{e}_{_{h_0}})\\
     & =  \mathbf{G}^q \left(  \mathbf{F}^r \, (\mathbf{e}_{_{l_0}} - \mathbf{e}_{_{h_0}}) \right)\\
\end{split}
\end{equation}

\noindent Let us also define the following vectors $\mathbf{u}_r$ for $r =  0, 1, 2, \cdots, k_0 - 1$: 
\begin{equation}
    \mathbf{u}_r = \mathbf{F}^r \, (\mathbf{e}_{_{l_0}} - \mathbf{e}_{_{h_0}})  \\
\end{equation}
and observe that: $\sum_{i=1}^n (\mathbf{u}_r)_{i} = 0$ for all $r =  0, 1, 2, \cdots, k_0 - 1$ because, as pointed out above, $\mathbf{F}$ conserves the sum of vector coordinates, and the sum of the coordinates of the initial vector $(\mathbf{e}_{_{l_0}} - \mathbf{e}_{_{h_0}})$ is equal to: $1 + (-1) = 0$.\\

\noindent Because $\mathbf{G}$ is a power of $\mathbf{F}$ and thus conserves the sum of vector coordinates,we can state that:
\begin{equation}
\begin{split}
\norm{\mathbf{F}^k \, (\mathbf{e}_{_{l_0}} - \mathbf{e}_{_{h_0}})} = \norm{\mathbf{G}^q \, \mathbf{u}_r} = \norm{\mathbf{G}(\mathbf{G}^{q-1} \, \mathbf{u}_r)} &  \leq (1 - g) \norm {\mathbf{G}^{q-1} \, \mathbf{u}_r} \\
 \norm {\mathbf{G}^{q-1} \, \mathbf{u}_r} & \leq (1 - g) \norm {\mathbf{G}^{q-2} \, \mathbf{u}_r} \\
\norm {\mathbf{G}^{q-2} \, \mathbf{u}_r} & \leq (1 - g) \norm {\mathbf{G}^{q-3} \, \mathbf{u}_r} \\
\vdots
\\
\norm {\mathbf{G}^{2} \, \mathbf{u}_r} & \leq (1 - g) \norm {\mathbf{G} \, \mathbf{u}_r} \\
\norm {\mathbf{G} \, \mathbf{u}_r} & \leq (1 - g) \norm {\mathbf{u}_r} \\
\end{split}
\end{equation}
In conclusion, 
\begin{equation}
\norm{\mathbf{G}^q \, \mathbf{u}_r} \leq (1 - g)^q \norm {\mathbf{u}_r} \\
\end{equation}
Let us call:
\begin{equation}
\beta = \sup\{ \norm{\mathbf{u}_0},\norm{\mathbf{u}_1},  \cdots, \norm{\mathbf{u}_{k_0 -1}} \}    
\end{equation}
and observe that:

\begin{equation}
\begin{split}
(1 - g)^q  = (1 - g)^{k/k_0} \, (1 - g)^{-r/k_0} &  \leq (1 - g)^{k/k_0} \, (1 - g)^{-(k_0 -1)/k_0} \\
(1 - g)^q & \leq \frac{\left((1 - g)^{1/k_0}\right)^k}{(1 - g)^{(1 -1/k_0)}} \\
\end{split}
\end{equation}
As  a  result:
\begin{equation}
\begin{split}
\norm{\mathbf{F}^k \, (\mathbf{e}_{_{l_0}} - \mathbf{e}_{_{h_0}})} = \norm{\mathbf{G}^q \, \mathbf{u}_r} &  \leq \frac{\left((1 - g)^{1/k_0}\right)^k}{(1 - g)^{(1 -1/k_0)}} \times \beta\\
\norm{\mathbf{F}^k \, (\mathbf{e}_{_{l_0}} - \mathbf{e}_{_{h_0}})} & \leq \left( \frac{\beta}{(1 - g)^{(1 -1/k_0)}} \right) \left((1 - g)^{1/k_0}\right)^k \\
\norm{\mathbf{F}^k \, (\mathbf{e}_{_{l_0}} - \mathbf{e}_{_{h_0}})} & \leq \left( \frac{\beta}{(1 - g)^{(1 - \mathcal{C})}} \right) \left((1 - g)^{\mathcal{C}}\right)^k \\
\end{split}
\end{equation}
or, equivalently, using equation \eqref{equa_22}: 

\begin{equation}
\norm{\mathbf{x}_{\epsilon}(t_0 +k) - \mathbf{x}(t_0 +k)} \leq \left( \frac{\epsilon \, \beta}{(1 - g)^{(1 - \mathcal{C})}} \right) \left((1 - g)^{\mathcal{C}}\right)^k \\
\end{equation}
\begin{equation}
\boxed{\norm{\mathbf{x}_{\epsilon}(t_0 +k) - \mathbf{x}(t_0 +k)} \leq \gamma_0 \left((1 - g)^{\mathcal{C}}\right)^k}
\end{equation}\\
where $\gamma_0 = \epsilon \, \beta/(1 - g)^{(1 - \mathcal{C})}$, $\mathcal{C}$ is the cohesiveness of the society, and $g$ is the overall generosity of the economy.\\

\noindent Because $(1 - g)^{\mathcal{C}} < 1$, the length of  $\mathbf{F}^k \, (\mathbf{e}_{_{l_0}} - \mathbf{e}_{_{h_0}})$ dwindles down to 0, as $k$ increases and, as a result:

\[ \lim_{k \rightarrow \infty}\mathbf{x}_{\epsilon}(t_0 +k) = \mathbf{x}(t_0 +k)\]
For values of $k$ that are large enough, we can state that:
\begin{equation}
    \mathbf{x}_{\epsilon}(t_0 +k)  = \mathbf{x}(t_0 +k) + \epsilon \, 
\mathbf{F}^k \, \left(\mathbf{e}_n - \mathbf{e}_1 \right)\approx \mathbf{x}(t_0 +k)      
\end{equation}

 The rate of convergence of $\mathbf{x}_{\epsilon}(t_0 +k)$ to $\mathbf{x}(t_0 +k)$ as $k$ increases is governed by $\left((1 - g)^{\mathcal{C}}\right)^k$, which is itself dependent on the cohesiveness $\mathcal{C}$ of the society and the overall generosity $g$ of the economy. The higher the cohesiveness and the overall generosity, the faster the wealthy agents recover the financial support they provide to the poor agents at the very bottom of the social hierarchy. With this result on the rate of convergence at hand, one may start thinking to promote the acts of generosity within society, and link this entire approach to religious precepts. It is not difficult, however, to realize that attempts to promote generosity for purely religious reasons, while the agents are exchanging lousy products in the economy, would lead to unsustainable socio-economic patterns. Like everything in nature, the generosity coefficient  $\alpha_i$ of the economy towards agent $i$ has a \textit{dual notion} associated with it, which is the degree to which the goods and services sold by agent $i$ are \textit{competitive} in the market. Consumers around the world have been willing to pay higher prices for Apple products, not because they want to be generous towards Apple, but because of the quality of the products they buy from it. The generosity coefficients are to be settled in the markets and any attempt to promote generosity for purely religious reasons would lead to low quality business exchanges, and fragile socio-economic systems. It is usually by considering both the concept and its dual that sustainability can be achieved and its analysis becomes effective. 
 
\section{Discussion and future perspectives}


The description of the income circulation model that was proposed in this paper was limited to closed economies, but if we decide to apply this model to the economy of the entire world, then this limitation becomes irrelevant. According to recent statistics (e.g., McKinsey \& Company, ``\textit{The 2020 McKinsey Global Payments Report}'', 2020), many countries around the world have witnessed a large increase in electronic payments over the last few years.  For example, electronic payments in 2020 exceeded 90\%, 75\% and 70\% in Sweden, UK, and US respectively. As a result, large portions of the data sets that are needed to compute the entries $f_{ij}$ in the world's income circulation matrix are available within the databases of the banks and credit card companies. The world's income circulation matrix would of course have billions of entries in it, but the computational infrastructures to process such large matrices do exist and have already been deployed for various applications such as the earlier versions of Google's PageRank algorithm. Using tools to identify the sub-graphs of the world's income circulation graph that are strongly connected and the ones that are not, one can spot the societies that are fragmented and the ones that are cohesive to help develop plans for supporting the poor and vulnerable according to the results outlined in the above sections. \\

To our knowledge, the income circulation model proposed in the preceding sections is the first one to cast the dynamics of business exchanges in linear algebraic equations such as equations \eqref{equ_9} and \eqref{equa_10}. There is no restriction on the model's time step $\Delta t = (t+1) -t$ which could theoretically be as small as $1$ second or millisecond, or even smaller than that. Hence, one would wonder whether we could think about the change of the wealth vector $\mathbf{x}$ over an infinitesimal interval $dt$ in an (imaginary) economy $\mathcal{E}_i$ where the agents are trading among each other in a \textit{continuous} fashion, as opposed to the discrete nature of the transactions we carry out in real-world economies. We could then attempt to express the difference:
 \[\mathbf{x}(t+dt) - \mathbf{x}(t)\]
 in terms of parameters and variables that characterize the business exchanges in this economy right at that very time instant $t$. A number of physicists and mathematicians have tackled such a question using the laws of statistical physics applied to a sufficiently large number of rational economic agents \cite{bouchaud, during, degong, degong*, yi}. It is our belief that, if reasonable rules are defined to characterize the \textit{continuous} nature of the trading among economic agents, the discrete equations \eqref{equ_9} and \eqref{equa_10} of income circulation would become, when reconsidered for the continuous case where the time step $\Delta t$ approaches 0, consistent with the kinetic and hydrodynamic-like models presented in \cite{bouchaud, during, degong, yi}. In particular, we argue that the \textit{economic process} through which money rises from the bottom tiers of a cohesive society to the top tiers, as a result of business exchanges (i.e., people paying for the goods and services which they need for their own survival and which tend to be owned by agents stacked up in the higher tiers $H$ of the society), is similar to the \textit{physics process} of convection that drives fluid molecules away from a heat source, which could be, for example, the surface of the ocean near the equator in atmospheric convection, or the lower heated plate in a Rayleigh–Bénard experiment. Both of these economic and physics processes are driven by some kind of a gradient.\\

There is, however, a clear distinction between physical systems and socio-economic ones in the way they handle the second half of the cycle, which gets triggered when the object of concern (molecules for the former systems, and money for the latter) reaches its destination at the end of first half of the cycle. More specifically, when the molecules that traveled far away from the heat source reach the cold source (that is the tropopause in the case of atmospheric convection, and the upper cold plate in the case of the Rayleigh–Bénard experiment), they become subject to another type of gradient that drive them back, statistically speaking, to where they came from \--- the heat source. As for money in socio-economic systems, as soon as it lands in the coffers of wealthy agents, it becomes subject to only one thing: the \textit{free will} of these agents. The $\epsilon$-support method we described in this paper comes down, in essence, to getting socio-economic systems to do to money what physical systems do to molecules in the second half of the convection cycle. If the top wealthy class of the society decides to deploy the $\epsilon$-support method to help the poorest individuals in the social hierarchy, then the cohesion of the society will be strengthened, and the trust among the agents will be reinforced. Otherwise, the society runs the risk to break down at the \textit{weakest} links of the income circulation graph, disintegrate into a fragmented society and trigger the class struggle dynamics we highlighted in relation to the structure of the income circulation matrix \eqref{equa_16}.

\begin{appendix}
\newpage
\appendixpage

\section*{\underline{Appendix A}: How the proposed income circulation model fits in the economics literature}\label{appA}

\vspace{1cm}

 The income circulation model we proposed in this paper (see equations \eqref{equ_9} and \eqref{equa_10}) falls under the topic of \textit{circular flow of economic activity}, which is usually presented in the form of a diagram called \textit{circular-flow diagram}. Historians of economic thought have traced the earliest ideas and concepts of economic circular flow back to 18th century economists John Law, Richard Cantillon and François Quesnay (see, for example, Antoin E. Murphy. ``John Law and Richard Cantillon on the circular flow of income'', in `Journal of the History of Economic Thought', 1993, pages 47-62), but the origins of the circular-flow diagram in its modern shape have been attributed to Frank Knight who described it in his booklet titled ``The Economic Organization'' (1933) as the ``the familiar figure of the `wheel of wealth'''. Later on, Paul A. Samuelson included the circular-flow diagrams in all editions of his textbook ``Economics'' from 1948 to 2010. In the more recent textbook ``Principles of Economics'' (8th edition) by N. Gregory Mankiw (2018), the circular-flow diagram is introduced as the first economic model to represent the organization of the economy. Then, to introduce the second economic model about the ``Production Possibilities Frontier'', the author indicates that most ``economic models, unlike the circular-flow diagram, are built using the tools of mathematics.'' The income circulation model outlined in equations \eqref{equ_9} and \eqref{equa_10} introduces the appropriate tools of mathematics to capture the dynamics that underlie the circular-flow diagram in the context of a closed economy where no credit is available to the economic agents. In more general terms, the financial accounting concepts of \textit{liability} and \textit{asset} were not considered at this stage in the model. A brief discussion of how this model can be scaled up to include governments, trade with other economies and financial institutions is presented in the next Appendix.

\newpage
\section*{\underline{Appendix B}: How the income circulation model can be scaled up}\label{appB}

\vspace{0cm}

 In an attempt to build the money circulation model (see  equations \eqref{equ_9} and \eqref{equa_10}) from the ground up using deductive reasoning, we omitted financial institutions, governments and international trade in the economy $\mathcal{E}$. It is, however, possible to scale up the proposed money circulation model, and include these three other aspects. Here is a brief summary of how the scale-up can take place:

\begin{enumerate}
    \item Governments can be included in the money circulation matrix $\mathbf{F}_{\tau}$ in the same way other agents (stores, supermarkets, corporations, employees, and so on) are represented in this matrix. The row entries corresponding to a government (municipal, provincial, state or national) would represent the taxes paid by other agents in the economy or, more specifically, the fractions of other agents' wealth paid as taxes to the government. Similarly, the government's column entries would refer  to the fractions of the government's ``wealth", which should be understood to refer to what we call the ``government's budget'' in economics literature, paid by the government to other agents (salaries for civil servants, social assistance paid individual citizens, subsidies, and so on). Finally, a government's diagonal entry would refer to what we call `budget surplus' or, in the case where financial institutions are included (see item 3 below), `deficit'. 
    
    \item International trade can be incorporated in the matrix model \eqref{equ_9} by introducing the concepts and tools of \textit{block matrices} in linear algebra. The global economy and the interactions among its agents can be captured in a large-scale block matrix, whose rows and columns refer to countries, the diagonal blocks represent money circulation matrices within the individual countries, and non-diagonal blocks capture the elements of international trade among countries.
    
    \item The introduction of banks and financial institutions in the economy gives rise to the variability of the monetary base $M$, and the possibility that the diagonal entries of the income circulation matrix $\mathbf{F}_{\tau}$ at time $\tau>0$ become negative, i.e., the agents are in debt, as opposed to saving money. Further research is needed to develop creative \textit{and} effective mathematical approaches to handle debt and, in more general terms, the question of liabilities and assets in the economy. One approach one can consider for such a purpose is to use an additive model to describe the economic activities in the society and capture the dynamics that underlie Frank Knight's `\textit{wheel of the wealth} (1933):
    
    \begin{equation}
      \begin{pmatrix}
          \text{model for income} & \\
          \text{circulation related} & \\
          \text{to the acts of} & \\
          \text{selling and buying} & \\
          \text{of goods and services}
      \end{pmatrix}  +         \begin{pmatrix}
          \text{model for the} & \\ 
          \text{economic activities} & \\
          \text{that involve the trading} & \\
          \text{of liabilities and assets}
      \end{pmatrix}
    \end{equation}

\vspace{0.5cm}

     The first term in the above sum refer to the income circulation model outlined in the equations \eqref{equ_9} and \eqref{equa_10} of the paper. The second term refer to a model that needs to be developed to handle assets and liabilities in society.
\end{enumerate}

\newpage

\section*{\underline{Appendix C}: List of the paths from an agent $i$ to another agent $j$ in the simple economy $\mathcal{E}_{ex}$ of 3 agents with the income circulation matrix $\mathbf{F}_{ex}$ specified in the paper:}\label{appC}

\begin{table}[h!]
    \centering
\begin{tabular}{ |c|p{6cm}|c| }
 \hline
  Pair $(i,j)$ of agents & \ \ \ \ Path from $i$ to $j$ & Length $k$ of the path from $i$ to $j$ \\
 \hline
 \hline
 (1,1) & $1 \rightarrow 2 \rightarrow 3 \rightarrow 1$  & 3 \\
       & $1 \rightarrow 2 \rightarrow 3 \rightarrow 1 \rightarrow 2 \rightarrow 3 \rightarrow 1$  & 6 \\
 \hline
 (1,2) &   $1 \rightarrow 2$  & 1 \\
       &   $1 \rightarrow 2 \rightarrow 3 \rightarrow 1 \rightarrow 2$  & 4 \\
 \hline
 (1,3) & $1 \rightarrow 2 \rightarrow 3$  & 2 \\
       & $1 \rightarrow 2 \rightarrow 3 \rightarrow 1 \rightarrow 2 \rightarrow 3$  & 5 \\
 \hline
 (2,1)    & $2 \rightarrow 3 \rightarrow 1$ & 2\\
         & $2 \rightarrow 3 \rightarrow 1 \rightarrow 2 \rightarrow 3 \rightarrow 1$ & 5\\
 \hline
 (2,2) & $2 \rightarrow 1 \rightarrow 3 \rightarrow 2$  & 3\\
    & $2 \rightarrow 1 \rightarrow 3 \rightarrow 2 \rightarrow 3 \rightarrow 1 \rightarrow 2 $  & 6\\
 \hline
 (2,3) & $2 \rightarrow 3$  & 1\\
  & $2 \rightarrow 3 \rightarrow 1 \rightarrow 2$  & 4\\
 \hline
 (3,1) & $3 \rightarrow 1$  & 1 \\
       & $3 \rightarrow 1 \rightarrow 2 \rightarrow 3 $  & 4 \\
 \hline
 (3,2) & $3 \rightarrow 1 \rightarrow 2 \rightarrow 3$  & 2\\
     & $3 \rightarrow 1 \rightarrow 2 \rightarrow 3 \rightarrow 1 \rightarrow 2 \rightarrow 3$  & 5\\
 \hline
 (3,3) & $3 \rightarrow 1 \rightarrow 2 \rightarrow 3$  & 3\\
      & $3 \rightarrow 1 \rightarrow 2 \rightarrow 3 \rightarrow 1 \rightarrow 2 \rightarrow 3$  & 6\\
 \hline
\end{tabular}    \caption{Paths to Connect Pairs of Agents in the Simple Economy $\mathcal{E}_{ex}$}
    \label{tab:1}
\end{table}

\newpage

\section*{\underline{Appendix D}: Proof of theorem \ref{theorem_1}}\label{appD}
\vspace{0.5cm}

 This theorem can be proven by induction on $k$, while also using the Perron–Frobenius theory for non-negative matrices (if need be, we can provide a detailed step by step proof for the theorem's statements). 

\newpage

\vspace{0.5cm}
\section*{\underline{Appendix E}: Proof of inequality:}\label{appE}
\vspace{0.2cm}
\begin{equation*}
    \norm{\mathbf{G}  \, \mathbf{u}} \leq (1 - g) \, \norm{\mathbf{u}} 
\end{equation*}
where:
\begin{enumerate}
    \item $\mathbf{G}$ is the positive column-stochastic matrix: 
    \[\mathbf{G} = \left[g_{ij} \right]_{\substack{i=1,n \\ j=1,n}} > \mathbf{0}_{n,n}\]
    \item for each $i \in \{1,2, \cdots, n\}$, $\alpha_i$ is the real number defined as: 
    \[ \alpha_i = \min_{j=1,n} g_{_{i,j}} >0 \]
    
    \item $g$ is the real number defined as:
    
    \[ g = \sum_{i=1}^n \alpha_i \in ]0,1[\]
    
    \item $\mathbf{u} = [u_1, u_2, \cdots , u_n]^T$ is an $n \times 1$ vector that satisfies the condition:
    
    \[ \sum_{j=1}^n u_j =0 \]
    
    \item the norm $\norm{\ }$ is the $l_1$ norm
    
\end{enumerate}

\vspace{0.5cm}

 \textbf{Note}: There are many different ways to approach the analysis of the linear-algebraic income circulation models introduced in this paper, thanks to the very rich body of knowledge that spans the algebra of nonnegative matrices, the Perron-Frobenius theory, graph theory and Markov chains. However, we tried our best to keep the mathematical treatment in the paper as simple as possible, with the intention to bring the physics aspects of socio-economic phenomena to the fore. For readers who are familiar with the mathematics of nonnegative matrices, we are using in the proof below the argument that a stochastic positive matrix induces a contraction mapping on the $n$-dimensional space spanned by wealth vectors and equipped with the $l_1$ norm.

\begin{proof}
\begin{equation}
\begin{split}
\norm{\mathbf{G}  \, \mathbf{u}} & = \sum_{i=1}^n \left|\sum_{j=1}^n g_{ij} u_j \right| = \sum_{i=1}^n \left|\sum_{j=1}^n (\alpha_i + (g_{ij}-\alpha_i)) u_j \right|  \ \ \ \  \ \ \ \  \left( \text{by definition of the } l_1 \text{ norm of } \mathbf{G}  \, \mathbf{u} \right) \\
 & = \sum_{i=1}^n \left|\sum_{j=1}^n (\alpha_i \, u_j + (g_{ij}-\alpha_i) \, u_j) \right| \\
 & = \sum_{i=1}^n \left|\, \alpha_i\left(\sum_{j=1}^n u_j\right) + \sum_{j=1}^n  (g_{ij}-\alpha_i) \, u_j \right|\\
 & = \sum_{i=1}^n \left| \sum_{j=1}^n (g_{ij}-\alpha_i) \, u_j \right| \ \ \ \ \ \ \ \ \ \ \ \ \ \ \ \ \ \ \ \ \ \ \ \ \ \left( \sum_{j=1}^n u_j = 0 \ \text{because of item 4 in the above list} \right)\\
 & \leq \sum_{i=1}^n \sum_{j=1}^n \left|(g_{ij}-\alpha_i) \, u_j \right| = \sum_{i=1}^n \sum_{j=1}^n (g_{ij}-\alpha_i) \, |u_j|  \ \ \ \ \ \ \left( (g_{_{i,j}} - \alpha_i)\geq 0 \text{ because }\alpha_i = \min_{j=1,n} g_{_{i,j}} \right) \\
  & \leq \sum_{i=1}^n \sum_{j=1}^n (g_{ij} |u_j|-\alpha_i |u_j|)= \sum_{j=1}^n \sum_{i=1}^n (g_{ij} |u_j|-\alpha_i |u_j|) \\
 & \leq \sum_{j=1}^n \left[ \left(\sum_{i=1}^n g_{ij} |u_j|\right )- \left( \sum_{i=1}^n \alpha_i |u_j|\right) \right]\\
& \leq \sum_{j=1}^n \left[ |u_j| \left( \sum_{i=1}^n g_{ij}\right) - |u_j| \left( \sum_{i=1}^n \alpha_i \right) \right] \ \ \ \ \ \ \ \ \ \ \left( \sum_{i=1}^n g_{ij} = 1 \text{\ because\ } \mathbf{G} \text{ is column-stochastic} \right)\\ 
& \leq \sum_{j=1}^n \left[ |u_j| - |u_j| \left( \sum_{i=1}^n \alpha_i \right) \right] = \sum_{j=1}^n \left[|u_j| \left[ 1 - \left( \sum_{i=1}^n \alpha_i \right) \right]  \right] \\ 
& \leq \left[ 1 - \left( \sum_{i=1}^n \alpha_i \right) \right]  \sum_{j=1}^n |u_j| = \left[ 1 - \left( \sum_{i=1}^n \alpha_i \right) \right]  \norm {\mathbf{u}} \\
\end{split}
\nonumber
\end{equation}
\vspace{-0.5cm}
\noindent In conclusion:
\begin{equation*}
    \norm{\mathbf{G}  \, \mathbf{u}}  \leq  (1- g) \norm {\mathbf{u}} 
\end{equation*}
where: \[ g = \left( \sum_{i=1}^n \alpha_i \right)\]. \\
\end{proof}



\end{appendix}

\newpage

\noindent \textbf{\Large About the authors:}
\\

\noindent AG and JH are two Canadian systems scientists with background in mathematics, physical chemistry and engineering science. They live in Toronto, Ontario, and can be reached at the following e-mail address:

\begin{center} \textit{ a2guerga@torontomu.ca}
\end{center}
Comments and feed-backs on this paper are welcomed and can be sent to the authors at the above e-mail address. \\

\newpage

\printbibliography

\end{document}